\documentclass{appolb}
\usepackage{graphicx}
\usepackage{amsmath,amssymb}

\begin{document}
\title{STRANGENESS IN QGP:\\ HADRONIZATION PRESSURE%
\thanks{Talk presented at the XXXI \ Max Born Symposium and HIC for FAIR Workshop\  ``Three Days \ of\ \ Critical Behavior \ in \ Hot and Dense QCD'', \  Wroclaw, \  Poland, June~14--16,~2013.}%
}
\author{Jan~Rafelski,  Michal~Petran%
\address{%
Department of Physics, The University of Arizona,Tucson, AZ 85721, USA\\[0.2cm]}
}%
\maketitle
\centerline{\date{Received February 14, 2014}}

\begin{abstract}
We review strangeness as signature of quark gluon plasma (QGP) and the hadronization process of a QGP fireball formed in relativistic heavy-ion collisions in the entire range of today accessible reaction energies. We discuss energy dependence of the statistical hadronization parameters within the context of fast QGP hadronization. We  find   that QGP breakup occurs for all energies at the universal  hadronization pressure $P = 80\pm 3\,\mathrm{MeV/fm}^3 $.
\end{abstract}

\section*{Preface}
The theoretical and later experimental exploration of the QGP phase of matter was one of the main research directions at CERN when in the early '80s  one of us (JR) met and worked with Krzysztof Redlich there. Krzysztof  came to CERN interested in the study of small thermodynamic systems and the hadron statistical bootstrap model of Hagedorn. This work lead him naturally to join the effort JR spearheaded to establish strangeness as signature of the new phase of hadronic matter, QGP.   The  challenges Krzysztof created for strangeness helped the physics case along, and helped to keep  research interest focused on strangeness for that long a time. While we two battled over physics issues, we developed a profound personal friendship. Kryzsztof,  Happy Birthday -- Jan!
\eject

\begin{figure}
\centering
\includegraphics[width=0.83\columnwidth]{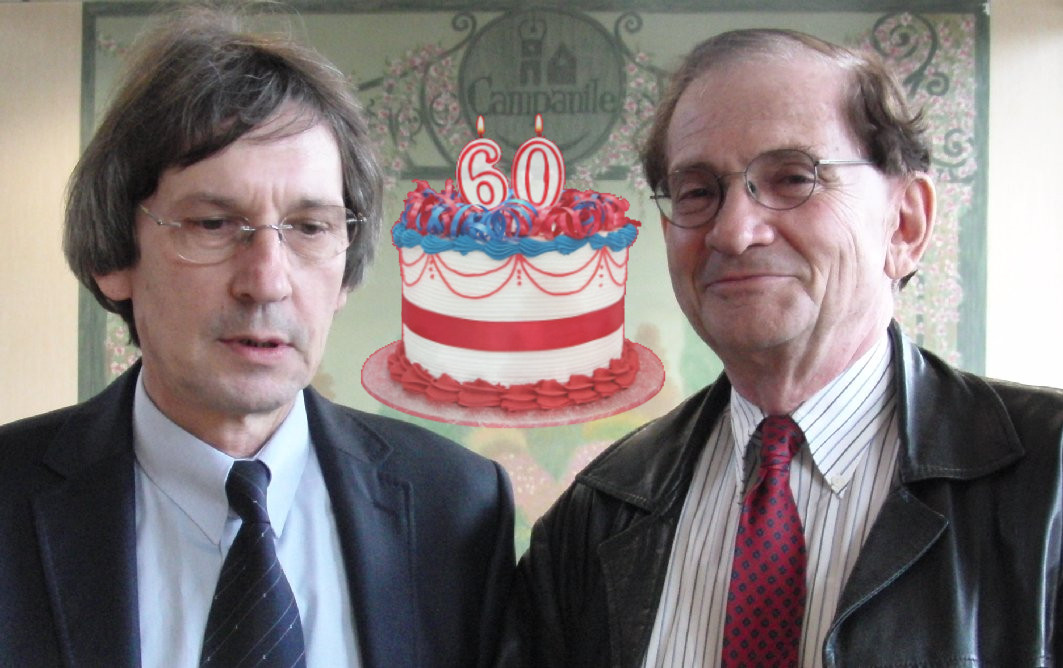}
\caption{\label{fig:Kryzsztof}Kryzsztof,  Happy Birthday -- Jan!}
\end{figure}

\section{Introduction}
We use the statistical hadronization model (SHM) to analyze the particle production. Our  aim is to learn about the properties of the deconfined quark-gluon plasma (QGP) phase and to study the dynamics of fireball evolution and  breakup, as related to the dynamics of particle freeze-out. We will show how such an analysis allows to conclude that the source of particles is indeed a QGP state, and how we can learn about physical properties of QGP. Our argument will be based on the earliest recognized signature of QGP,  strange particle production; for an early review see Ref.\cite{Rafelski:1982ii}. It is very important to keep in mind that the effects we consider as a signature of QGP are in general of magnitude 1.5--3, and thus depend on proper model implementation and a careful analysis. 

The need for precise and detailed study of QGP and its strangeness signature  has been recognized  early on by Krzysztof Redlich~\cite{Redlich:1985zg}. For the past 30 years one of us (JR) executed an approach that took  Krzysztof guidance into account. It is for this very reason that we never study the $pp$ reaction system, or use it as a reference to develop a result. In   a small $pp$ system the outcome of the experiment  depends on how the events evolve in terms of parton scattering and how the final state is selected by triggers. These and  many  other features  distinguish  $pp$ from the large $AA$ nuclear collision system and comparing these two is risque. 

When considering reactions between large nuclei $A$--$A$, for purpose of this discussion one can think of the case $A=$Pb or Au. In such most central collisions about $2\times 2,000$  partons (primordial quarks or/and gluons) interact. Only such relativistic heavy ion collision experiments provide  us the opportunity to create and study in a wide range of centrality and reaction energy a new state of matter -- QGP. The asymmetric case, $pA$ collisions are of interest but are far more difficult to interpret in our scheme and the data are here not as available. All said this report addresses solely $AA$ reaction. 

We consider here particle production  from CERN-SPS, BNL-RHIC to LHC, thus  an energy range that changes by  nearly 3 orders of magnitude from  $\sqrt{s_{NN}}=6.26$\,GeV (equivalent to  lowest SPS case of $20A\,\mathrm{GeV}$) to $\sqrt{s_{NN}}=2\,760$\,GeV at LHC today. Moreover, a wide range of geometric collision centrality is considered in these experiments with variation in participant number by more than an order of magnitude. As a consequence,  the resulting  range in particle production yields spans 5 orders of magnitude. 

We will  briefly review strangeness as signature of QGP in Secs. \ref{hadroniz} and \ref{sSsec}. In subsection \ref{XiPhi} we will describe in qualitative terms a simple multistrange particle yield ratio signature  that suggest chemical nonequilibrium scenario in hadronization of quark-gluon plasma (QGP). We then describe in  subsection \ref{sSphases} a   measure of strangeness abundance, the ratio of strangeness to entropy produced which allows to count the ratio of strange to all degrees of freedom in the QGP phase

This is followed  in section \ref{param} by a discussion of values of SHM parameters across all reaction energies. Of most interest are parameters which  being nearly constant connect across energies a common feature of the QGP fireball. We show that hadronization occurs for most systems after a swing-in for low reaction energy at $\gamma_q\simeq 1.6$ and  $T \simeq 140\,\mathrm{MeV}$. 

We study the bulk physical properties of the fireball in Section~\ref{bulk} We begin  by characterizing in subsection \ref{volume} the fireball volume. In subsection \ref{bulkP} we find adding up partial pressure contributions of all particles (both directly observed and inferred in the fit procedure) a hadronization pressure  $\sum_i P_i\equiv P = 80\pm 3\,\mathrm{MeV/fm}^3 $. This value  we find in all but  very few reaction systems available for analysis:  The exception we show here is when the collision  energy decreases to $20\,\mathrm{AGeV}$, equivalent to $ \sqrt{s_{NN}} =6.26 \, \mathrm{GeV}$ in central Pb--Pb collisions at SPS and/or most peripheral collisions  (not shown here) at the  higher RHIC and LHC reaction energy. We close in subsection \ref{sSsec} with a discussion of the values $s/S$ that are obtained in the SHM analysis.

We close this short report discussing our key findings and  presenting a general outlook in section  \ref{conclude}. The considerable difference in the outcome of SHM analysis occurring between two lowest energies studied at SPS: $\sqrt{s_{NN}}=6.26$ and $7.61$ GeV implies a threshold of new physics   between these reaction energies. This  result shows an opportunity for the Beam Energy Scan (BES) program at RHIC to identify the onset of quark deconfinement via the study of hadron  multiplicity yields.  

\section{Examples of strangeness signatures  of QGP}\label{hadroniz}
\subsection{Multistrange hadrons}\label{XiPhi}
In the abstract of Ref.\cite{Rafelski:1982ii} a particularly important diagnostic role is attributed to the possible overpopulation of the phase space occupancy of strangeness as measured among hadrons. This is today  characterized by the parameter $\gamma_s$ which surpasses the chemical hadron phase equilibrium at  $\gamma_s=1$.  As described in 1982, considering  the high QGP strangeness yield,  and a fast breakup of chemically equilibrated QGP fireball, we can expect a large $\gamma_s\simeq 2$.  Such anomalous yield is a signature of high strangeness abundance state being the source of particles. The possibility that $\gamma_i>1$ for both light and strange quarks is  indicating both a QGP source and its fast explosive breakup. It is  not surprising that the only SHM variant  which describes the experimental results at LHC consistently as a function of centrality is the  chemical \emph{non-equilibrium} SHM where both strangeness $\gamma_s>1$ and   light quark  $\gamma_q > 1$ are allowed in the particle yield fit process~\cite{Petran:2013lja}.
 
As the above discussion shows, it is important to distinguish between the strangeness phase space occupancy in the QGP $\gamma_s^Q$, and $\gamma_s^H$ which we measure observing final state  hadrons. This distinction is again part of the careful consideration of how strangeness enhancement operates in different phases~\cite{Redlich:1985zg}.  In general we omit the superscript `H' and so when a quantity without superscript appears it addresses the hadron phase space. These two phases are different; in the QGP we evaluate the population of free quarks and gluons, as compared to counting constituent strange quarks in the rather massive hadrons. 

At LHC   we find $\gamma_s^H = 2.0$ corresponds to $\gamma_s^Q=0.88$ in the QGP~\cite{Petran:2013dea}. On the other hand a value $\gamma_s^Q\to 1$ is seen at RHIC. We should note that considering a very large reaction volume at LHC the overall yield of strangeness is greatly enhanced. However, a 10-15\% reduction of $\gamma_s^Q$ and thus of expected strangeness abundance is clearly present indicating a very interesting and novel development in study of QGP properties at LHC energies. This effect is just the opposite to the expectation that at LHC overpopulation of the QGP strangeness should occur as high strangeness yield produced early on is preserved to the end of QGP evolution~\cite{Letessier:2006wn}. 

\begin{figure}
\centering
\includegraphics[width=0.98\columnwidth]{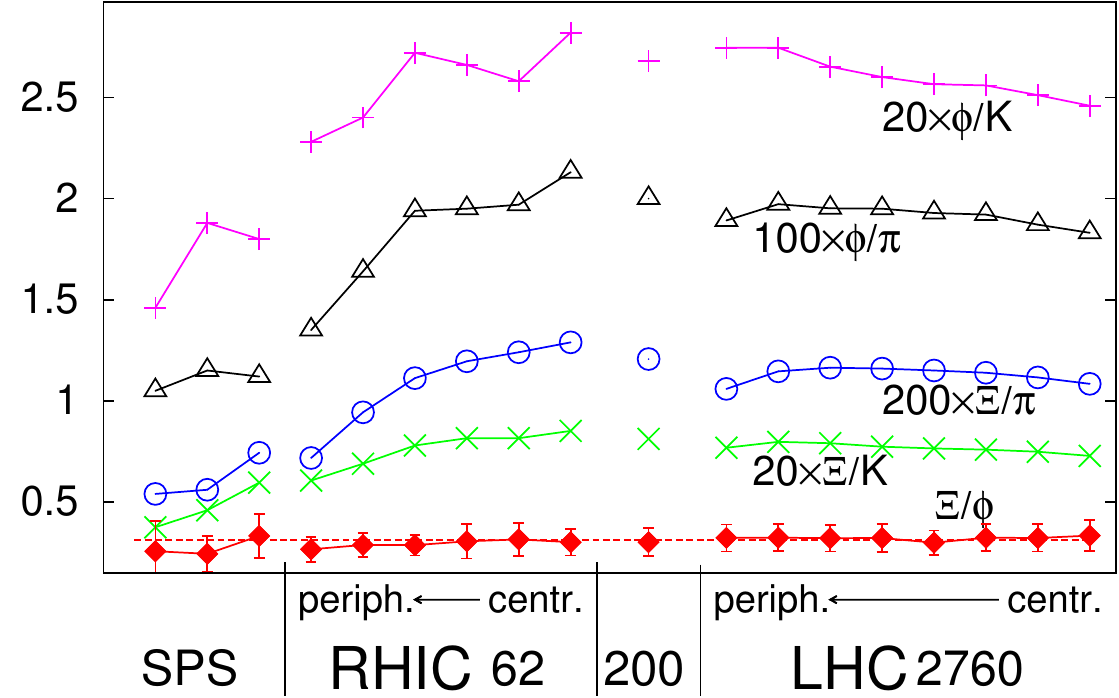}
\caption{\label{fig:ratios} Ratios of multistrange  particle yields (bottom, $\Xi/\phi$) compared to ratios with varying strangeness content schematically as function of both centrality and system energy, both rising to the right.}
\end{figure}

The special role of multistrange hadrons   $\Xi(qss)$,$\phi(s\bar s)$  (and $\Omega$) has  been noted in Ref.\cite{Rafelski:1982ii}. As the energy rises from SPS to LHC we see by inspecting the bottom line in Figure~\ref{fig:ratios} that $\Xi/\pi$  ratio remains constant. This behavior imposes a constraint on properties of the source of these particles across all energies and centralities and barring a  coincidence, it implies that the fireball properties other than $\gamma_s$ which cancel  remain  constant -- note that in Figure~\ref{fig:ratios}  we only show the three top SPS energies and omit the low energy result.

This implied constancy of fireball properties in particular must apply to  the light quark phase space occupancy $\gamma_q$ and the freeze-out temperature $T$ necessary to describe this ratio~\cite{Petran:2009dc,Petran:2013dea}. We further note  $\Xi/K$ and $\phi/K$ ratios seen in Figure~\ref{fig:ratios}. These are (nearly) proportional to $\gamma_s$. We see that these ratios measured  at different collision energies and centralities vary by a factor of two, implying a similar change in $\gamma_s$. Confirming $\gamma_s$ as the source of the ratio variations  are the two $\gamma_s^2$-dependent ratios  $\Xi/\pi$ and $\phi/\pi$, which vary yet more strongly. Variation of $\gamma_s$ is consistent with the proposed approach to chemical equilibrium of strangeness abundance in the QGP fireball.  Depending on how large, and hot the fireball is,   more strangeness can be produced:  An increase of initial temperature with reaction energy, and an increase in volume with reaction centrality helps to increase fireball strangeness yield.

\subsection{Counting degrees of freedom in the fireball}\label{sSphases}
To establish the presence of free quarks in the fireball based on particle yield data, we compare the strange quark degree of freedom  to all degrees of freedom. For this purpose we evaluate the ratio of strangeness density (or yield) to entropy density (or yield) $s/S|^Q$. Especially if hadronization is fast there can be little change both in entropy and strangeness yield or even density. Thus the ratio $s/S|^Q$ has the superscript `Q" to indicate that its value as  measured with hadrons is (nearly) the same as  in the QGP phase. Furthermore,  this ratio is up to factor $\simeq 4$, the entropy per relativistic particle, is  the ratio of strange to all other  active degrees of freedom. When counting strangeness  we include $s\bar{s}$  pairs hidden in  $\eta,\eta^\prime,\phi$.

\begin{figure}
\centering
\includegraphics[width=0.9\columnwidth]{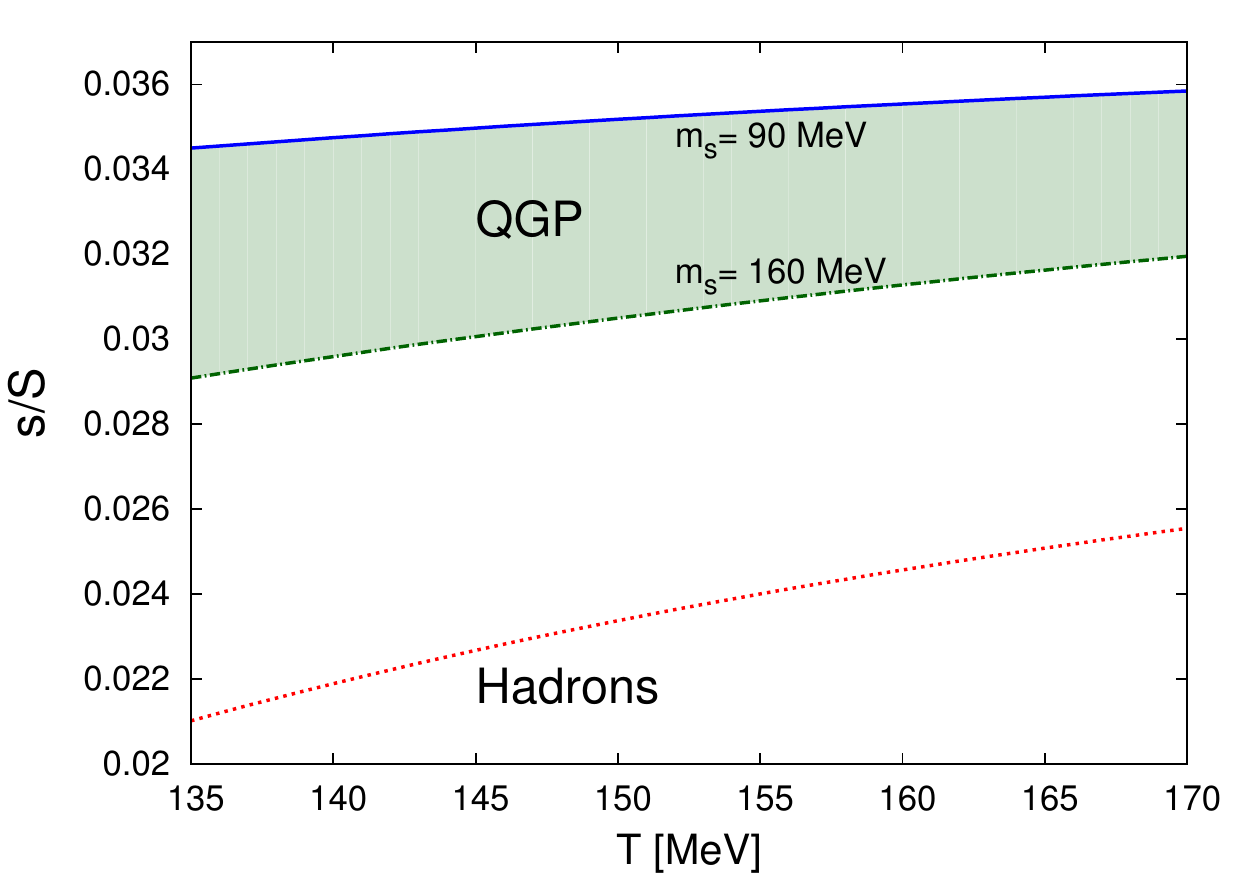}
\caption{\label{fig:sOverS}Strangeness over entropy in QGP, domain braced up by  $m_s=90\,\mathrm{MeV}/c^2$ (solid blue line), and bottom $m_s=160\,\textrm{MeV}/c^2$ (dash-dotted green line), and for the hadron gas at the very bottom of figure (dashed red line). }
\end{figure}

For a chemically equilibrated QGP comprising free quarks and gluons, this ratio is:
\begin{equation}
\left.\frac{s}{S}\right|^Q\simeq \frac{1}{4}\frac{n_s}{n_s + n_{\bar{s}} + n_q + n_{\bar{q}} + n_G   } 
= \frac{\frac{g_s}{2\pi^2}T^3(m_s/T)^2 K_2(m_s/T)}{(g2\pi^2/45)T^3+ (g_s n_f/6)\mu^2_q T}
\simeq \frac{1}{35}.
\end{equation}
This ratio increases with $\mathcal{O}(\alpha_s)$ interactions to $s/S \to 1/31= 0.0323$. In chemical non-equilibrium QGP, this result will be multiplied by the QGP strangeness phase space occupancy $\gamma_s^Q$. Time evolution of strangeness, entropy and the phase space occupancy can be computed using the kinetic theory described in  Ref.~\cite{Rafelski:1982ii,Letessier:2006wn}. 

A benchmark values of $s/S|^H$ is obtained for chemical equilibrium  of hadrons; that is  $\gamma_s,\gamma_q=1$. This hadron phase property is shown as the bottom line in  Figure~\ref{fig:sOverS}, as a function of temperature. We note that the benchmark value is about  $s/S|^H\simeq 0.023$.  If we multiply this value by a factor 1.5 we find ourselves in midst of the QGP ratio in  Figure~\ref{fig:sOverS}, where the large value domain arises   due to  uncertainty in the strange quark mass - within 1s.d. $90\le m_s\le 160$\,MeV. The central value of this range is the reference strange quark mass corresponding to the scale of temperature here of relevance.  Clearly measuring  the ratio $s/S$ and for completeness $T$ experimentally we can argue if the source was QGP, or a hadron gas~\cite{Kuznetsova:2006bh}. 

Even though  hadron phase $s/S$ is ``only'' a factor $1.5$ below the QGP value, these are very different and   distinguishable scenarios provided we can achieve the necessary precision in data analysis, which is the case in the chemical non-equilibrium SHM.  An increase in strangeness yield by 50\% is in principle easily visible  bearing in mind that the experimental uncertainty in study of particle yields  is in general  at, and below, 10\%.

\section{SHM parameters of the hadronizing fireball}\label{param}
\begin{figure}
 \centerline{\includegraphics[width=0.5\columnwidth]{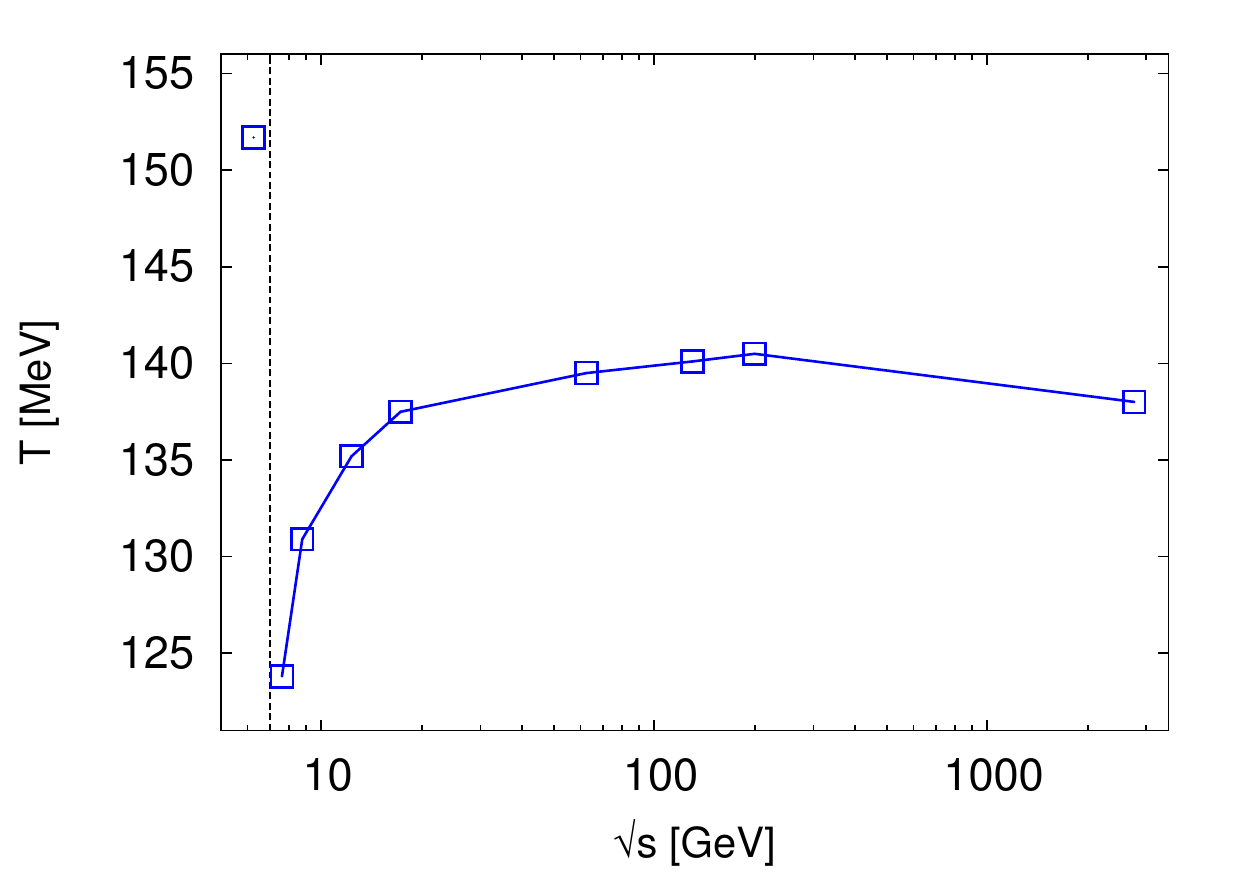} 
\includegraphics[width=0.5\columnwidth]{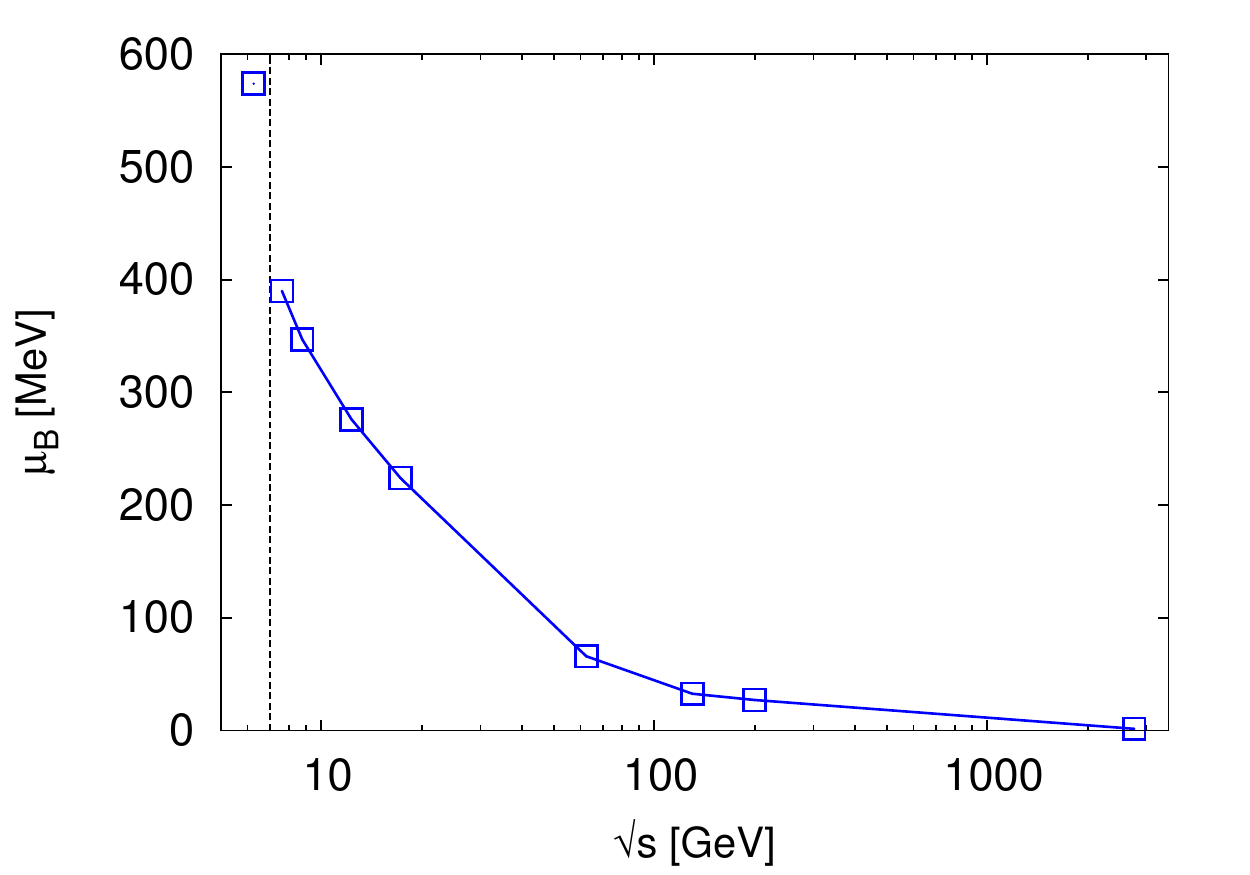}}
\caption{\label{fig:mub}Chemical freeze-out temperature $T$ (on left) and baryochemical potential $\mu_B$ (on right)  as a function of CM collision energy, obtained in non-equilibrium SHM for the most central collisions. The lowest available $\sqrt{s_{NN}}=6.26\,\mathrm{GeV}$ result is  separated off by a vertical dotted line.}
\end{figure}

For all considered systems, the  freeze-out temperature  creeps at high energy towards  $T\simeq 140\,\mathrm{MeV}$; see Figure~\ref{fig:mub}  on left. Omitting the low energy value,  $T$ changes overall by about 10\%, seen from the highest value at $T\simeq 140$\,MeV.  On the other hand,   baryochemical potential $mu_B$ depicted in Figure~\ref{fig:mub} on right decreases rapidly with  collision energy. It varies by factor 500, from being a fraction of the nucleon mass, to MeV range for LHC energies. This shows that the fireball of quark matter changes considerably in its compsition, from being baryon-rich  to being baryon-free. Note that where $T$ is relatively small, $\mu_B$ is largest.   We will see  that this effect compensates the effect that the low value of temperature has on pressure and it is the high baryon density at low SPS energies that allows us to find the remarkable constancy of   hadronization pressure $P$.

\begin{figure}
 \centerline{\includegraphics[width=0.5\columnwidth]{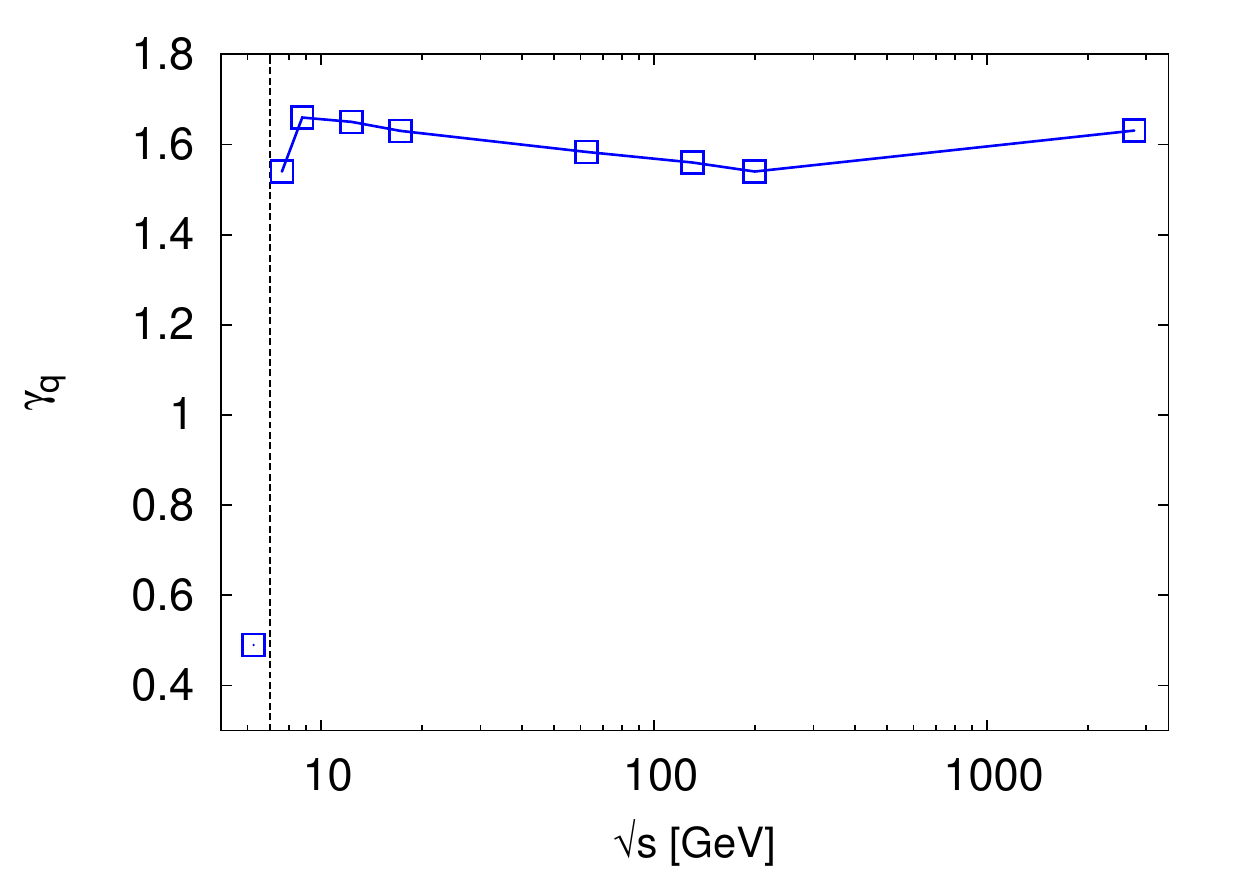}
\includegraphics[width=0.5\columnwidth]{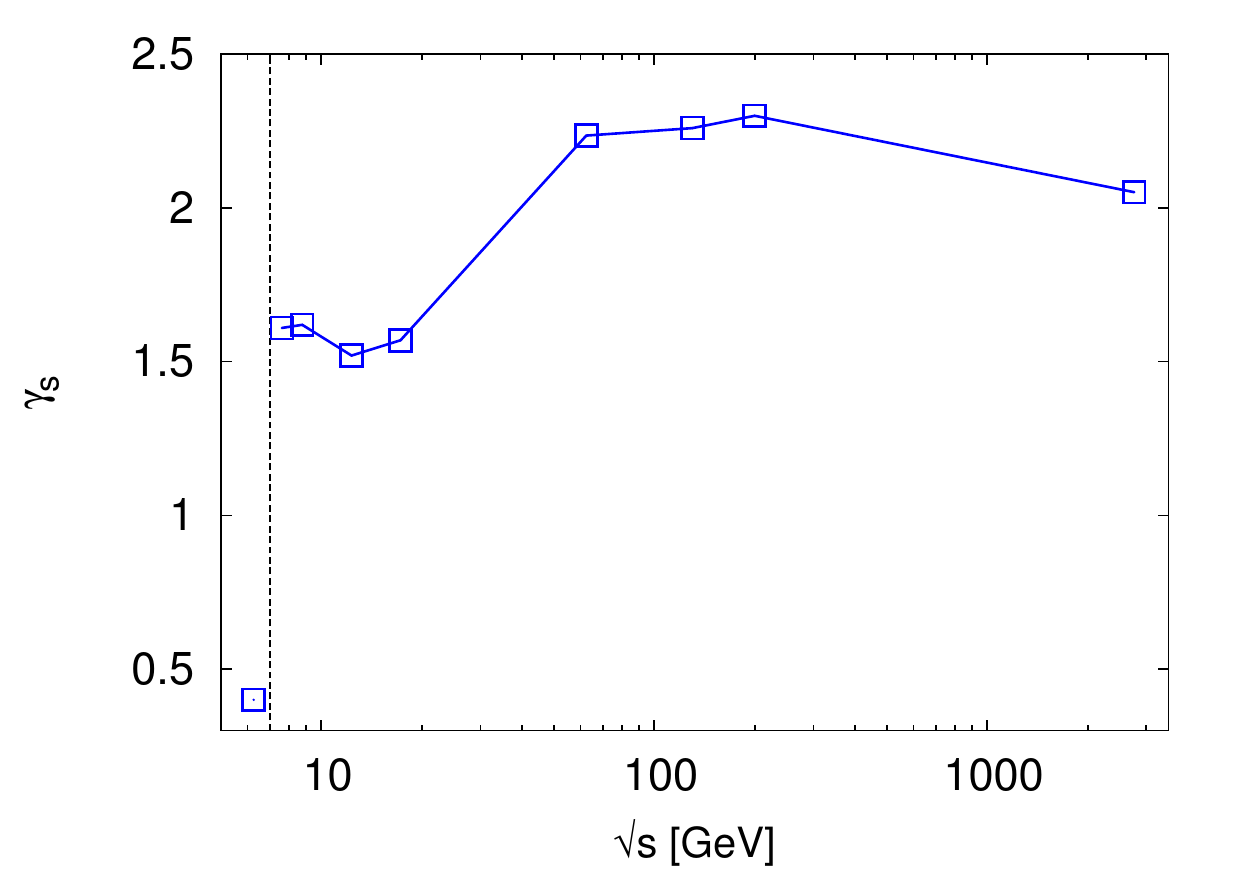}}
\caption{\label{fig:gammmas}Phase space occupancies: $\gamma_s$ for strangeness and $\gamma_q$ for the light quarks.}
\end{figure}

The analysis of the precise Pb-Pb collision data from LHC allowed us to rule out SHM variants other than the chemical non-equilibrium~\cite{Petran:2013lja}. This justifies using the non-equilibrium SHM approach at SPS and RHIC. In fact we find a clear preference for  phase space occupancy  $\gamma_q\simeq 1.6$ across all energies, as shown on left in Figure~\ref{fig:gammmas}. The exception is   the lowest SPS energy. Thus in most cases  the light quark phase space is overpopulated as is expected for   hadronization of the high entropy QGP state~\cite{Kuznetsova:2006bh}.

The right panel of Figure~\ref{fig:gammmas} shows the strange phase space occupancy $\gamma_s$. There is a more significant variation than for $\gamma_q$ seen in left panel. This is so since it is harder in scattering processes to reach QGP strangeness equilibrium as compared to the saturation of entropy abundance. Strangeness phase space is  $1.5 < \gamma_s < 2.4$ depending on the reaction energy,  excluding from this discussion the lowest SPS energy data set at $\sqrt{s_{NN}}=6.26\,\mathrm{GeV}$ as this reaction energy is clearly within a different class of events.

The lowest energy Pb-Pb collisions at SPS, $20\,\mathrm{AGeV}$ (i.e. $\sqrt{s_{NN}}=6.26\,\mathrm{GeV}$), stands out in terms of these and in fact all other statistical parameters. Higher hadronization temperature $T\sim 152\,\mathrm{MeV}$ and phase space occupancies $\gamma_{s,q}<1$ indicate that the scenario of sudden hadronization of QGP may not apply in this energy region and calls for further experimental effort in the future.

\section{Bulk properties of the fireball}\label{bulk}
\subsection{Hadronization volume of QGP}\label{volume}
\begin{figure}
\centering
\includegraphics[width=0.9\columnwidth]{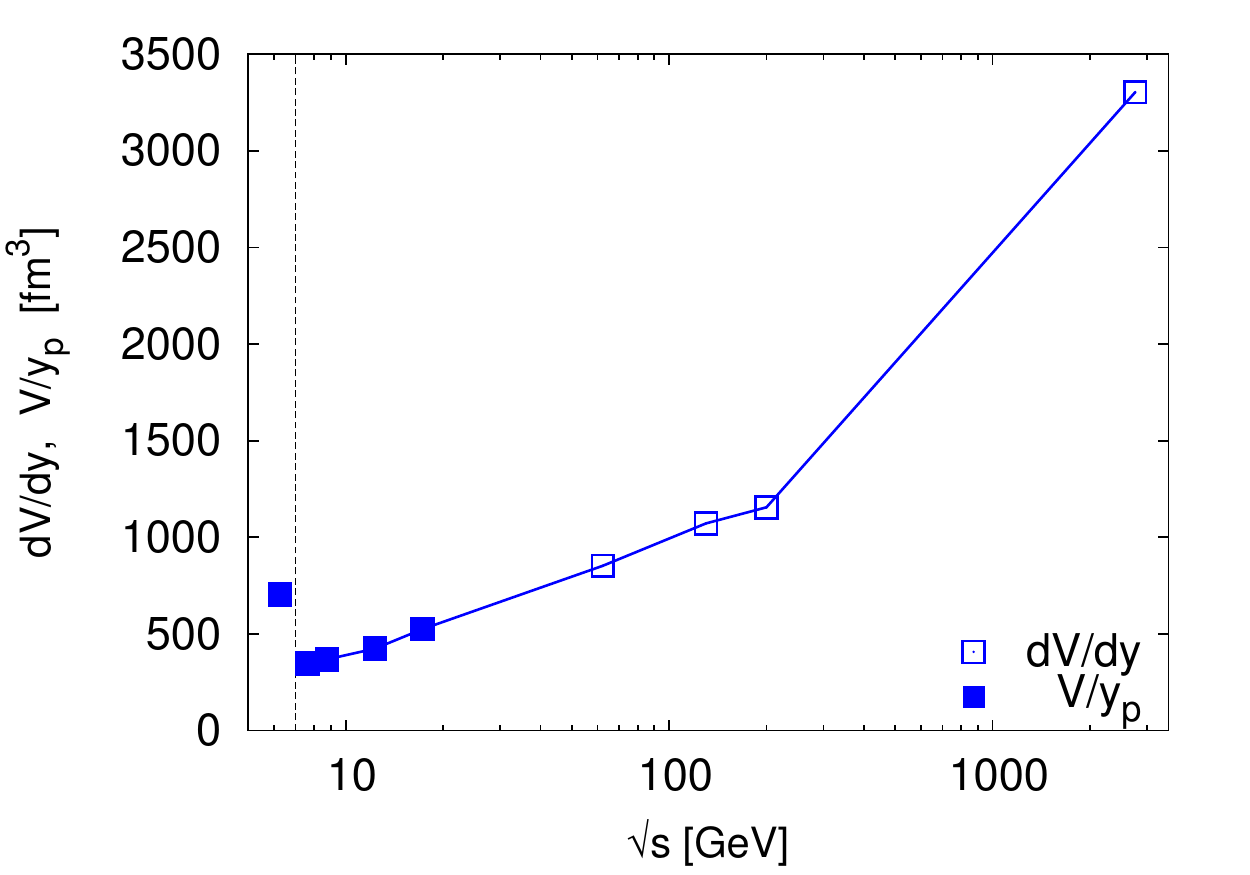}
\caption{\label{fig:vol} $dV/dy$   per unit of rapidity at LHC, RHIC  (open symbols) and $V/y_p$   for SPS (full symbols). See text for further details.}
\end{figure}

The main variation in the property of the fireball as the energy and centrality changes is the associated fireball volume parameter. In other words the change in hadron yields which spans five orders of magnitude is predominantly originating in the size of the fireball at time of its breakup. The overall particle yields is normalized by  the fireball volume $V$ in fixed target experiments at SPS,  and by the distribution $dV/dy$ at RHIC and LHC where only the most central rapidity bin $y\in (-0.5, 0.5)$ is available. In order to relate the two measurements we recall that the normalization volume for a fixed target SPS experiment can be also written as
\begin{equation}
V = \int\limits_{0-\delta y}^{y_{p}+\delta y} dy \left(\frac{dV}{dy}\right),\qquad 
\left.\frac{dV}{dy}\right|_{y_p/2} \le \frac{V}{y_p}
\end{equation}
where $y_p$ is the projectile rapidity in the lab frame, and  $\delta y$ is the small and natural single particle distribution spread which is clearly visible in experimental data. ${dV}/{dy}$ has been  is some experiments also evaluated and presented at SPS energies. In Figure~\ref{fig:vol}, along with $dV/dy|_{y=0}$  for collider experiments, we show $V/y_p$ for fixed target experiments. The lowest energy SPS point  aside, separated by vertical line, we see that the equivalent central rapidity reaction volume grows with energy and this trend accelerates in the LHC domain.

\subsection{Bulk hadronization pressure}\label{bulkP}
\begin{figure}
\centering
\includegraphics[width=0.9\columnwidth]{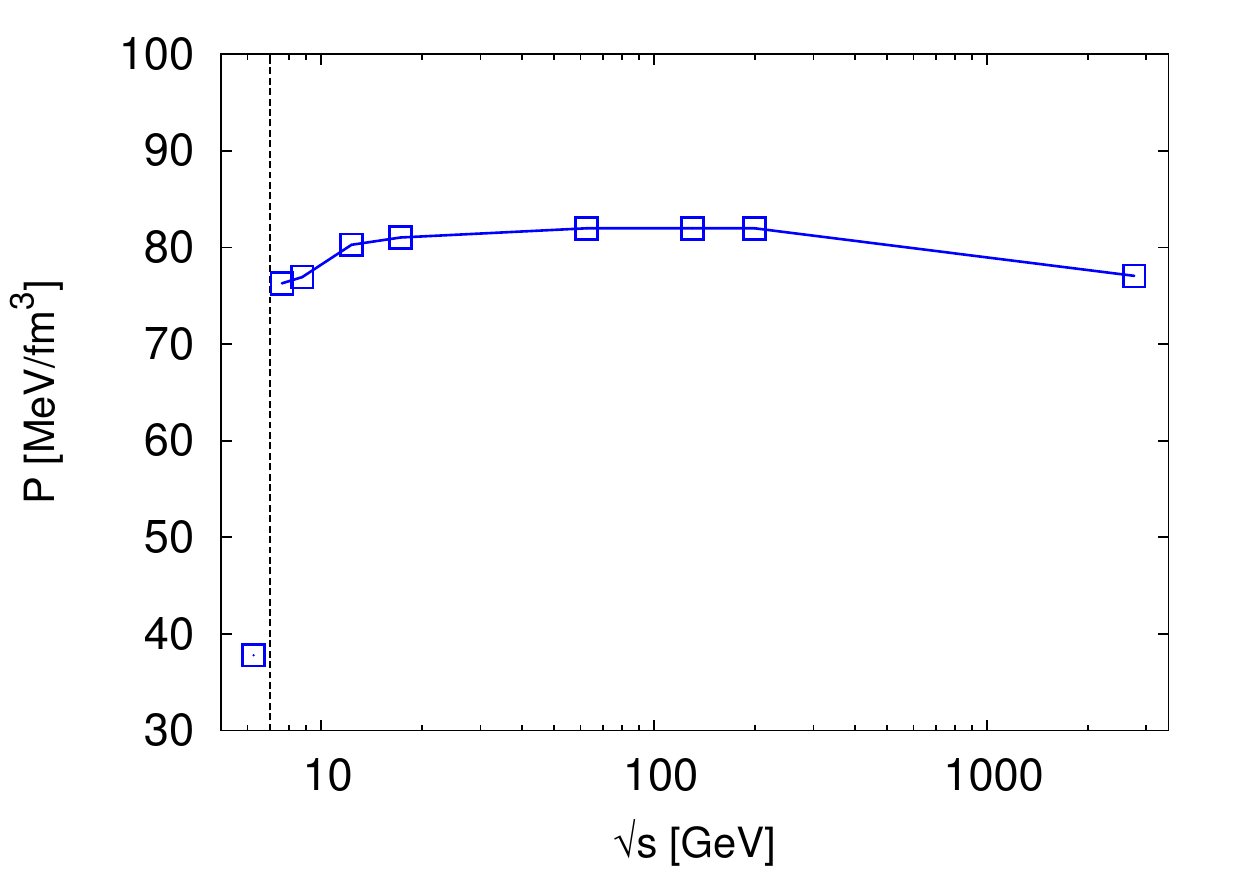}
\caption{\label{fig:pressure}Hadronization pressure for most central collisions as a function of collision energy.}
\end{figure}

We evaluate the pressure of the hadronizing QGP fireball by adding up partial gas component contribution of all emerging hadron fractions, obtained for both particles experimentally measured, and (as yet) unobserved hadrons. The result for the pressure $P$ across all SPS, RHIC, LHC energies is seen Figure~\ref{fig:pressure}.  We find  $P = 80\pm 3\,\mathrm{MeV/fm}^3$. Similarly to prior discussion, hadronization pressure shows a qualitative difference for the lowest SPS energy data set at $\sqrt{s_{NN}}=6.26\,\mathrm{GeV}$.The key result is that hadronization occurs for all other energies at the same critical pressure to within $\pm 4\%$.  The Universality of  hadronization pressure was originally proposed based on SPS data analysis in~\cite{Rafelski:2009jr}.

This result implies dynamical fine tuning since the pressure in hadrons is originating in many different components, and the contributions shift as we move from baryon rich fireball at low energy to a nearly baryonfreelimit at LHC.  The pressure of final state hadrons  corresponds according to Gibbs phase equilibrium considerations to the pressure of quarks, gluons, at hadronization. These nearly massless particles contribute a  radiation pressure  $P_{\rm rad}\propto g_{eff}(T)T^4$, where $g_{eff}(T)$ refers to the degeneracy and number of particle species contributing to pressure. The high power of $T$ means that the 10\% changes in $T$  is compensated by variation of other properties of the fireball at hadronization, in particular baryon and strangeness abundance. 

We thus see that  fits to particle production data over a wide range of energy (and centrality, see~\cite{Petran:2013lja}) produce the same universal hadronization pressure. In our opinion this is the most surprising  result of the present analysis of particle production in heavy ion collisions. We here recall that Figure~\ref{fig:vol} shows the overall normalization $dV/dy$ (and $V/\Delta y$ equivalent for SPS) dependence on the collision energy, which changes by a factor of 7, and some SHM parameters vary even more widely, see e.g. $\mu_B$ depicted on right in the Figure \ref{fig:mub}.  Full consideration of hadronization conditions as a function of collision energy $\sqrt{s}$ and centrality in the light of this result will follow.

\subsection{Strangeness specific abundance}\label{sSsec}
\begin{figure}
\centering
\includegraphics[width=0.9\columnwidth]{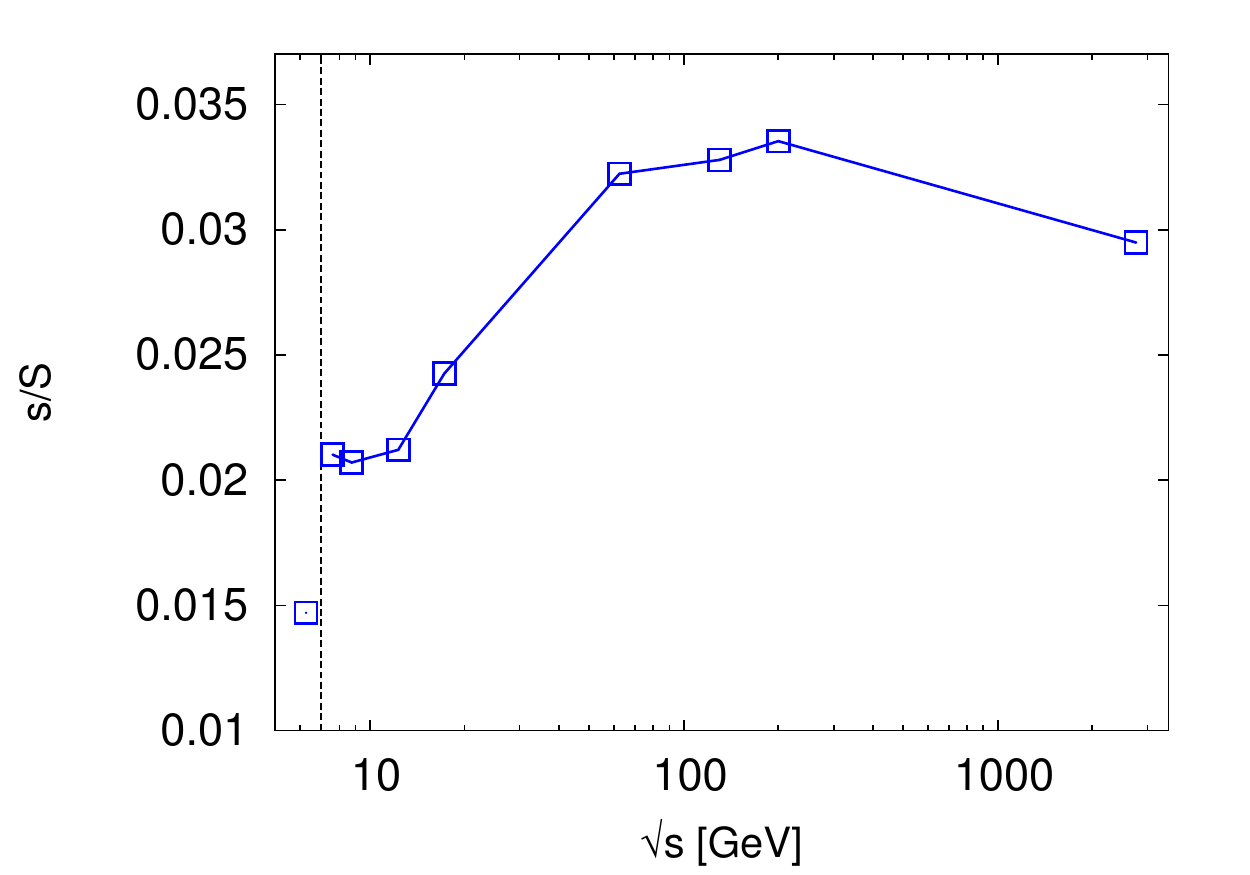}
\caption{\label{fig:sOverSdata}Strangeness over entropy $s/S$ following from particle yield analysis as function of $\sqrt{s_{NN}}$.}
\end{figure}

The outcome of our particle yield analysis at SPS, RHIC and LHC  for the most central collisions for the ratio $s/S$  is shown in Figure~\ref{fig:sOverSdata}. We note that the highest energy LHC value near 0.030 is noticeably lower as compared to the three RHIC points clustering at the value 0.0325. Both these results are in fact within error bar  compatible with each other and imply nearly strangeness equilibrated QGP phase, compare Figure~\ref{fig:sOverS}.

The LHC value is the most precise result and it is definitively somewhat lower than the average of the three RHIC results. Furthermore, we see  as energy is reduced from the RHIC to SPS that the value of $s/S$ decreases. We interpret this as evidence that QGP formed at SPS is below strangeness chemical equilibrium yield $\gamma_s^Q< 1$. However, based only on the $s/S$ result one could also argue that at SPS we have   nearly equilibrated HG phase. The consideration of other bulk properties, in particular of hadronization pressure $P$ indicates  common hadronization behavior and this favors the argument that all but the lowest SPS energy produces also QGP. Similar interpretation follows from the study of the ratio $\Xi/\phi$ as reported above. 

We also see that $s/S$ increases in a the very small interval  $\sqrt{s_{NN}}=(6.26,7.61)$\,GeV by 50\% as the energy is increased from the lowest SPS energy point $\sqrt{s_{NN}}=6.26$ to the next. This indicates  that the lowest $20A$GeV SPS projectile reaction energy belongs to another class of events. We interpret this as threshold of deconfinement.

\section{Remarks and conclusions}\label{conclude}
\subsection{Remarks}
This report is an update of ongoing research which addresses the universality of hadronization conditions and as such we emphasizes the relation  to our earlier work and do not discuss work by other groups. The here presented results have been obtained using the fully updated SHARE with CHARM SHM program~\cite{Petran:2013dva}. This program is all but simple. It comprises about  $ 7500$ lines of code. In its setup our program largely outperforms other comparable programs in terms of physics analysis capabilities. 

For example, SHARE is the only SHM implementation, which fits the precise LHC hadron yield data at the time of writing this proceedings. Results from SPS and RHIC has been refitted using the updated code (with only weak decay pattern taken over) and the results from previous publications~\cite{Rafelski:2009jr,Rafelski:2004dp,Letessier:2005qe,Rafelski:2009gu} were thus checked. Our code can use both the data on particle multiplicity as well as bulk physical properties of the fireball to obtain unmeasured particle yields and thus it has considerable predictive power. 

\subsection{Conclusion}
Given the precise LHC particle production data we have been able to demonstrate that chemical nonequilibrium is present in the final hadron state. This is consistent with our consideration of  the sudden QGP hadronization leading to  $\gamma_{s,q}>1$~\cite{Petran:2013lja,Petran:2013qla}.  While the light quark occupancy is almost constant approaching the critical pion condensation value of $\gamma_q \to 1.6$, the strangeness phase space occupancy  $\gamma_s$ varies as a function of energy, and centrality. For all systems in which we believe to see QGP we find $\gamma_s>1$.
 
We find  the same universal hadronization condition of the QGP fireball, the critical pressure of $P=80\pm3\,\mathrm{MeV/fm}^3$. Considering the large difference in the initial conditions (collision energy and centrality), the constancy of hadronization pressure is a strong evidence for universal hadronization conditions of QGP fireball created in heavy-ion collisions. This universal property of hadronization $P\simeq $Const. validates  the sudden  mechanism of hadron production. In the sudden hadronization model  all hadrons are produced directly from QGP, and their abundance does not further evolve in ensuing rescattering. This means that in a rescattering study of   hadron kinetic freeze-out one would find the hadron chemical freeze-out temperature $T_k$  above the temperature $T$ of QGP breakup which we obtained in the fit and which is  the temperature governing the hadron abundances. 

The lowest SPS energy considered, $\sqrt{s_{NN}}=6.26\,\mathrm{GeV}$, is found to be different from all higher energy as witnessed by vastly different fit outcome in terms of both statistical model parameters and bulk properties. The particle production in this system  has a clearly different pattern and could originate in a confined hadron fireball. The qualitative difference in the fitted SHM parameter values as well as in the hadronization pressure implies a different physical environment. The beam energy range of $20-30\,\mathrm{AGeV}$, that is center of mass $\sqrt{s_{NN}} = 6.26-7.61\,\mathrm{GeV}$ is the most interesting region in study of   the onset of QGP creation and hadronization. The Beam Energy Scan  program   at RHIC~\cite{Sumbera:2013kd} and forthcoming experimental facilities, such as NICA~\cite{Kekelidze:2013nla}, and FAIR-GSI-CBM Experiment~\cite{Friman:2011zz}, target this particular collision energy domain to explore the  onset of deconfinement. Our analysis shows that this study of  strange particle production can lead to interesting results. 

\section*{Acknowledgment}
This work has been supported by a grant from the U.S. Department of Energy, grant DE-FG02-04ER41318.

\end{document}